\documentclass[twocolumn,tighten]{aastex6}

\usepackage{placeins}
\usepackage{natbib}
\usepackage{amsmath}
\usepackage{color}
\usepackage{enumerate}

\begin{document}

\newcommand\be{\begin{equation}}
\newcommand\en{\end{equation}}

\shorttitle{Planet-driven Spiral Arms: Implications} 
\shortauthors{Bae \& Zhu}

\title{PLANET-DRIVEN SPIRAL ARMS IN PROTOPLANETARY DISKS: II. IMPLICATIONS}

\author{Jaehan Bae\altaffilmark{1,2,3} and Zhaohuan Zhu\altaffilmark{4}}

\altaffiltext{1}{Department of Terrestrial Magnetism, Carnegie Institution for Science, 5241 Broad Branch Road, NW, Washington, DC 20015, USA} 
\altaffiltext{2}{Department of Astronomy, University of Michigan, 1085 S. University Avenue, Ann Arbor, MI 48109, USA} 
\altaffiltext{3}{Rubin Fellow}
\altaffiltext{4}{Department of Physics and Astronomy, University of Nevada, Las Vegas, 4505 South Maryland Parkway, Las Vegas, NV 89154, USA}

\email{jbae@carnegiescience.edu}

\begin{abstract}
We examine whether various characteristics of planet-driven spiral arms can be used to constrain the masses of unseen planets and their positions within their disks.
By carrying out two-dimensional hydrodynamic simulations varying planet mass and disk gas temperature, we find that a larger number of spiral arms form with a smaller planet mass and a lower disk temperature.
A planet excites two or more spiral arms interior to its orbit for a range of disk temperature characterized by the disk aspect ratio $0.04\leq(h/r)_p\leq0.15$, whereas exterior to a planet's orbit multiple spiral arms can form only in cold disks with $(h/r)_p \lesssim 0.06$.
Constraining the planet mass with the pitch angle of spiral arms requires accurate disk temperature measurements that might be challenging even with ALMA.
However, the property that the pitch angle of planet-driven spiral arms decreases away from the planet can be a powerful diagnostic to determine whether the planet is located interior or exterior to the observed spirals. 
The arm-to-arm separations increase as a function of planet mass, consistent with previous studies; however, the exact slope depends on disk temperature as well as the radial location where the arm-to-arm separations are measured.
We apply these diagnostics to the spiral arms seen in MWC~758 and Elias~2--27.
As shown in \citet{bae17}, planet-driven spiral arms can create concentric rings and gaps, which can produce more dominant observable signature than spiral arms under certain circumstances. 
We discuss the observability of planet-driven spiral arms versus rings and gaps.
\end{abstract}


\section{INTRODUCTION}
\label{sec:introduction}

Recent observations with state-of-the-art telescopes have imaged multi-armed spirals in protoplanetary disks (e.g., MWC~758, \citealt{grady13}, \citealt{benisty15}, \citealt{reggiani17}; SAO~206462, \citealt{muto12}, \citealt{garufi13}, \citealt{stolker16}, \citealt{maire17}; Elias~2--27, \citealt{perez16}; AB Aur, \citealt{tang17}).
While the origin of the observed spiral arms is not clearly understood, one compelling possibility is gravitational interaction between a (proto)planet and the underlying disk.
If the observed spiral arms are indeed launched by a planetary companion, they can provide a unique opportunity to gain crucial insights into their formation and coevolution with their host disks.

While there have been many recent efforts to explain the observed spiral arms with planetary companions, different numerical simulations do not always agree.
For example, it is debatable whether a potential planet has to be located interior or exterior to the observed spiral arms.
Also, the number of planets needed to explain the observed spiral arms is not clear: if we detect two spiral arms, is one planet enough to explain both arms, or do we need a second planet?

In the companion paper (\citealt{baezhu17}, hereafter Paper I), we described the mechanism by which a planet excites multiple spiral arms in the underlying protoplanetary disk.
Building on this understanding, in the present paper we carry out a suite of two-dimensional hydrodynamic simulations and investigate how characteristics of planet-driven spiral arms, such as the number of spiral arms, pitch angle of spiral arms, arm-to-arm separation, and the relative strength of spiral arms, vary as a function of disk temperature and planet mass.
The main aim of the parameter study is to examine whether such characteristics of observed spiral arms can be used to constrain the mass and/or position of yet unseen planet.

The observability of planet-driven spiral arms is determined mainly by their openness and the magnitude of the perturbations they produce:  the more opened a spiral arm is and the larger perturbation a spiral arm produces, it will be more readily observable.
Based on three-dimensional hydrodynamic calculations and radiative transfer simulations, \citet{dongfung17} suggested that at least a Saturn-mass planet is required to excite detectable spiral arms with current observational capabilities.
Spiral arms driven by smaller-mass planets may not be directly detectable.
However, they still can create observable signatures: concentric rings and gaps.
In \citet{bae17} we showed that each spiral arm launched by a planet can create its own gap through shock dissipation.
This means that when a planet excites multiple spiral arms, it can create multiple gaps in the disk.
Pressure maxima (i.e., rings) can develop between those gaps, potentially trapping solid particles.
Preferentially in disks with a low viscosity, it is possible that low-mass planets not capable of generating observable spiral arms can still induce sufficient trapping of solid particles that can be observable \citep[][see also Section \ref{sec:solar_nebula} of the present paper]{bae17}.
The generation of multiple rings and gaps by planet-driven spiral arms might also have implications for terrestrial body assembly in the solar nebula, as discussed later.

This paper is organized as follows.
In Section \ref{sec:formation}, we briefly summarize the formation mechanism of planet-driven spiral arms discussed in detail in Paper I.
In Section \ref{sec:characteristics}, using a suite of two-dimensional hydrodynamic simulations with various planet mass and disk temperature, we examine whether characteristics of observed spiral arms can be used to constrain the mass and/or position of an unseen planet. 
In Section \ref{sec:application}, we apply the diagnostics to the spiral arms observed in the disks around MWC~758 and Elias~2--27, and discuss their potential origin(s).
In Section \ref{sec:gaps}, we discuss when spiral arms would be more readily observable than rings and gaps, and vice versa, and discuss the potential role of pressure bumps created by Jupiter's core in the solar nebula.
We summarize our findings in Section \ref{sec:summary}.

\begin{figure*}
\centering
\epsscale{1.15}
\plotone{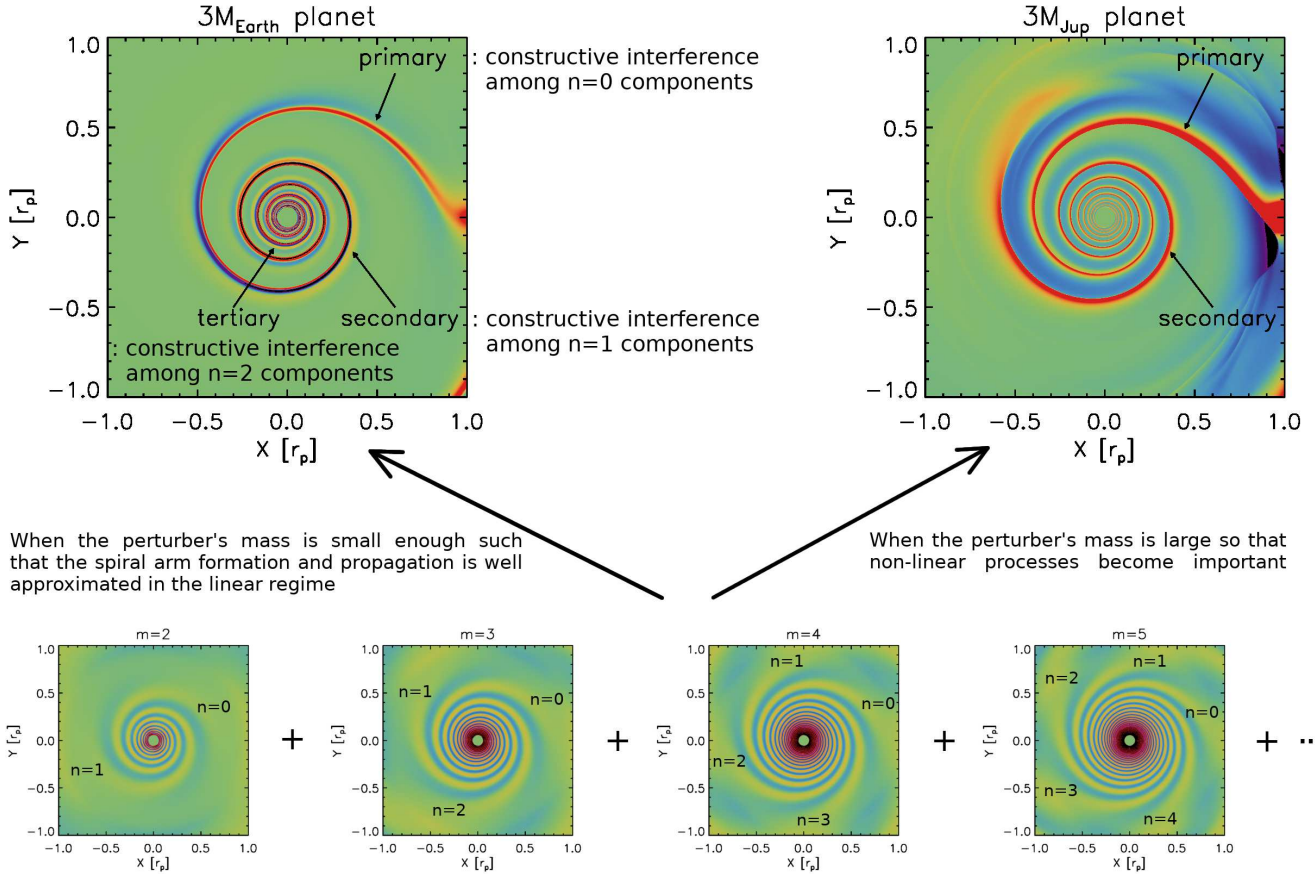}
\caption{An illustration of the formation of multiple spiral arms by a planet. In the upper panels, we present the disk surface density distributions from our two-dimensional hydrodynamic simulations with (left) a 3 Earth-mass planet and (right) a 3 Jupiter-mass planet orbiting around a solar-mass star. The central star is fixed at $(X,Y)=(0,0)$ and the planets are located at $(X,Y)=(1,0)$. The lower panels present the disk surface density distributions obtained with single Fourier-decomposed azimuthal modes (i.e., $m=2, 3, 4, 5, $...) whose magnitudes are chosen to be consistent with the perturbation driven by a 3 Earth-mass planet when added up (see Section 3.1 of Paper I). Note that the $m$th azimuthal mode excites $m$ wave modes, which are labeled with $n=0, 1, ..., m-1$. When the perturbations driven by single Fourier azimuthal modes are superimposed on to each other, they create coherent structures (i.e., spiral arms) as seen in the upper left panel. It is $n=0$, $n=1$, and $n=2$ components from each azimuthal mode that form the primary, secondary, and tertiary arms, respectively. As the planet mass grows non-linear effects become important, which affect to the pitch angle of spiral arms, number of spiral arms, separation between spiral arms, etc.}
\label{fig:sch}
\end{figure*}

\section{PLANET-DRIVEN SPIRAL ARM FORMATION MECHANISM}
\label{sec:formation}

The gravitational potential of a planet can be decomposed into a Fourier series, a sum of individual azimuthal modes having azimuthal wavenumbers $m=0, 1, 2, ..., \infty$.  
Through the resonance between the rotation of the planet's potential and the epicyclic motion of disk material, the $m$th Fourier component of the potential launches $m$ wave modes at its Lindblad resonances \citep[][see also the review by \citealt{shu16}]{goldreich78a,goldreich78,goldreich79}.
In the lower panels of Figure \ref{fig:sch} we present as examples the spiral wave patterns excited by individual azimuthal modes with $m=2,3,4,$ and 5.

When the mass of a planet is small enough so that non-linear effects can be safely ignored, we can simply add up the perturbations driven by individual azimuthal modes to reconstruct the perturbation driven by the full potential of the planet.
The upper left panel of Figure \ref{fig:sch} shows the resulting density perturbation in the disk.
In this example three spiral arms are launched interior to the planet's orbit: the primary arm forms right at the vicinity of the planet, whereas additional arms, denoted as secondary and tertiary, start to appear at a distance (in radius) from the planet.
In the Fourier representation, it is $n=0$, $n=1$, and $n=2$ components from individual azimuthal modes having different $m$ that generate the primary, secondary, and tertiary arms.
As can be seen in the lower panels of Figure \ref{fig:sch}, non-zero $n$th components excite out of phase initially; however, the propagation of wave modes depend on the azimuthal wavenumber in a way that they can be in phase as they propagate.
Exterior to the planet's orbit, the $n=0$ components generate the primary arm and the $n=m-1$ components generate the secondary arm.

As the planet mass increases, non-linear effects become increasingly important.
One of the main outcomes is that spiral arms become more opened, because spiral arms from a larger mass planet produce stronger shocks so propagate at faster speeds. 
Varying planet mass can thus have influence on the pitch angle of spiral arms, arm-to-arm separation, but also the total number of spiral arms excited by a planet, as can be inferred from the upper right panel of Figure \ref{fig:sch}.

\subsection{A Generalized Analytic Formula for the Phases of Spiral Arms}
\label{sec:phase_equation}

An important feature in the planet-driven spiral arm formation mechanism is that the formation of both primary and additional arms can be understood as a linear process when the planet mass is sufficiently small.
We can thus make use of linear wave theory to predict phases of spiral arms.
Here we provide a generalized analytic formula that can be used to fit observed spiral arms or to mimic spiral arms in models without carrying out planet-disk interaction simulations.

For low-mass planets (e.g., $\lesssim$ gap-opening mass), the following phase equation provides the phases of both primary and additional arms as a function of radius:
\begin{eqnarray}
\label{eqn:phase_m3}
\phi_{m,n} (r) = &-&{\rm sgn}(r - r_p) { \pi \over 4m} + 2\pi  {n \over m}  
\nonumber\\
&-& \int_{r_m^\pm}^{r} {\Omega(r') \over c_s(r')} \left| \left(1- {r'^{3/2} \over {r_p^{3/2}}} \right)^2 - {1 \over m^2} \right|^{1/2} {\rm d}r'.
\end{eqnarray}
Here, $r_p$ is the radius of the planet's circular orbit, $r_m^{\pm}  = (1 \pm 1/m)^{2/3}~r_p$ is the outer (with plus sign) and inner (with minus sign) Lindblad resonances, $m$ is the azimuthal wavenumber, $n=0,1,...,m-1$ represents individual wave modes, $\Omega$ is the disk rotational frequency, $c_s$ is the sound speed, and $r_p$ is the radius of the planet's circular orbit.
As we showed in Paper I, phases of spiral arms follow the dominating azimuthal mode with $m \approx (1/2) (h/r)_p^{-1}$ well in the linear regime, where $(h/r)_p$ is the disk aspect ratio at planet's orbit.
The primary arm phase can be calculated with $m=2\pi(h/r)_p$ and $n=0$.
As can be inferred from Equation (\ref{eqn:phase_m3}), one can simply shift the primary arm phase by $2\pi(n/m)$ in azimuth for additional arms, where $n=1$ and 2 for the secondary and the tertiary arms in the inner disk and $n=m-1$ for the secondary arm in the outer disk.
The derivation of Equation (\ref{eqn:phase_m3}) can be found in Section 2 of Paper I.

\begin{figure*}
\centering
\epsscale{1.15}
\plotone{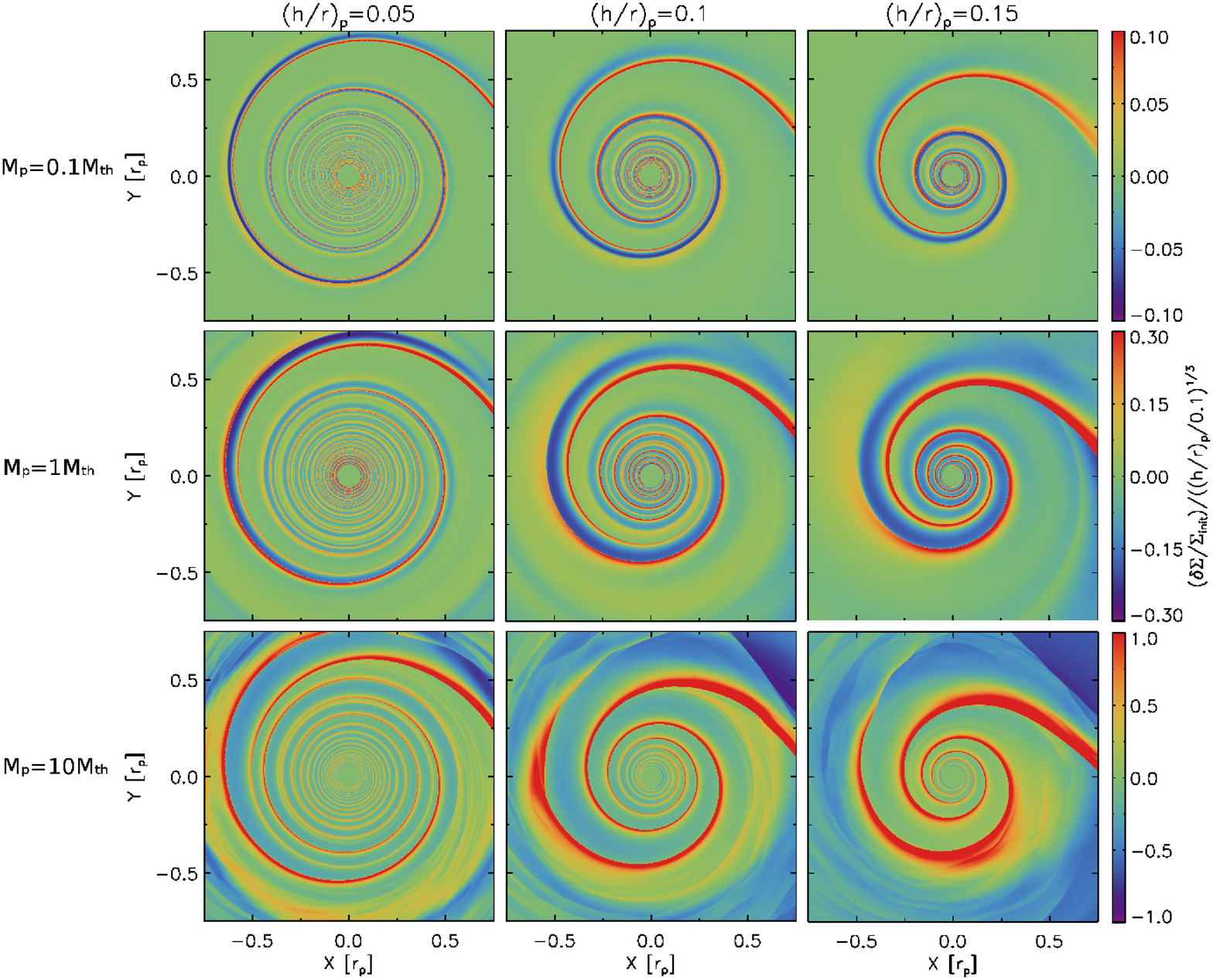}
\caption{Two-dimensional distributions of the perturbed surface density $\delta\Sigma/\Sigma_{\rm init}$ for various disk aspect ratio values $(h/r)_p$ and planet masses $M_p$, where $\delta\Sigma = \Sigma-\Sigma_{\rm init}$ and $\Sigma_{\rm init}$ denotes the initial surface density. The color scheme is adjusted by a factor of $((h/r)_p/0.1)^{1/3}$ such that planet-driven spiral arms with different $(h/r)_p$ values produce a similar level of perturbation. The planet is located at $(X,Y)=(1,0)$.}
\label{fig:mosaic}
\end{figure*}

\begin{figure*}
\centering
\epsscale{1.15}
\plotone{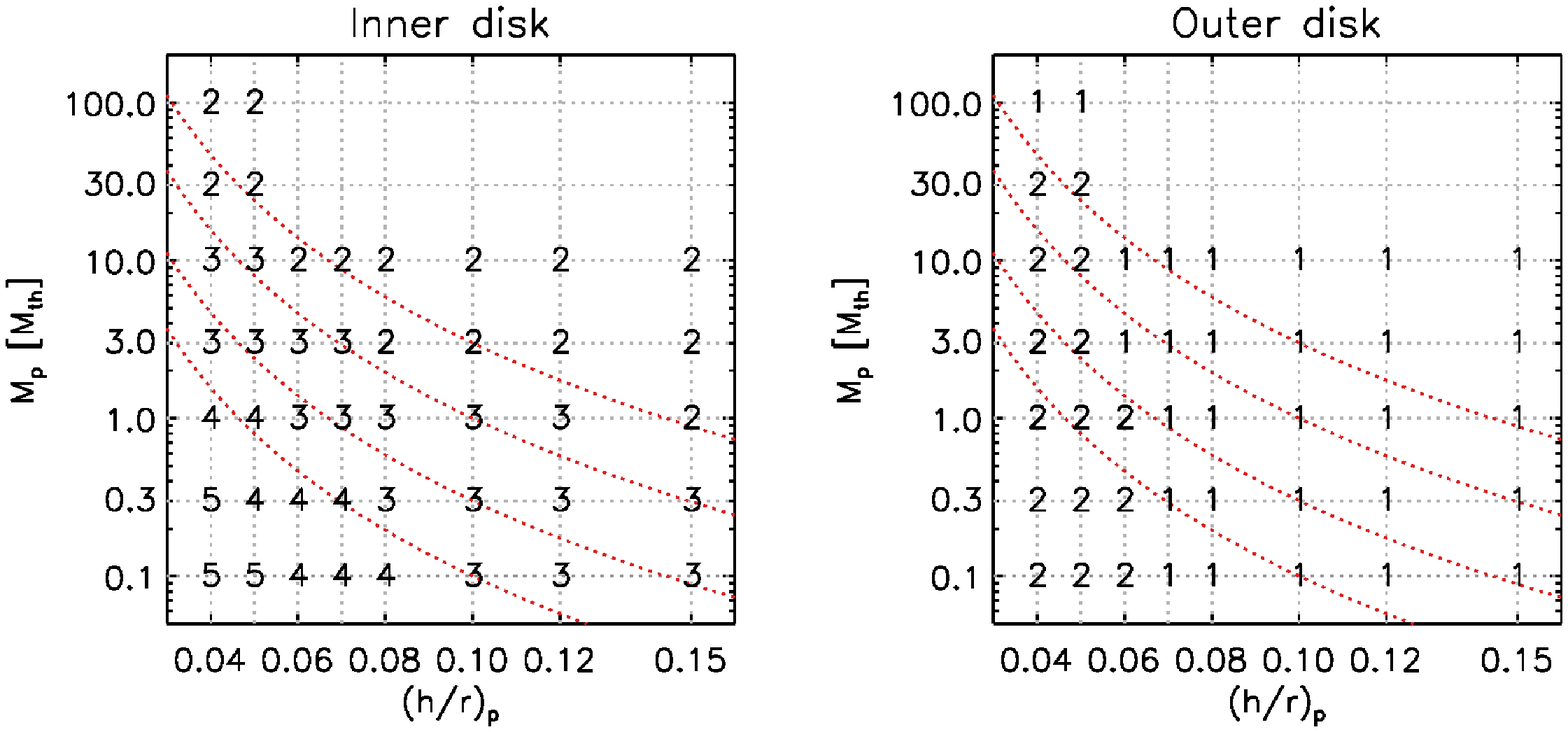}
\caption{Number of spiral arms with various values of $(h/r)_p$ and $M_p$ in the (left) inner disk ($0.2~r_p \leq r \leq 1~r_p$) and (right) outer disk ($1~r_p \leq r \leq 5~r_p$). In general, a larger number of spiral arms form for a lower mass planet in a disk with a smaller $(h/r)_p$ value. 
The numbers presented in this plot is the total number of spiral arms formed within the entire radial region defined above. We find spiral arms can merge as they propagate inward so the number of spiral arms at a given radius can be smaller than the numbers presented here. Merging of spiral arms preferentially occurs in cold disks with $(h/r)_p < 0.1$. See bottom left panel of Figure \ref{fig:mosaic} for example, in which case the primary and secondary arms merge and only two spiral arms are left at $r \lesssim 0.25~r_p$.  The red dotted curves present a constant planet-to-star mass ratio of $M_p/M_*= 3\times10^{-3}, 10^{-3}, 3\times10^{-4},$ and $10^{-4}$, from top to bottom. For a solar-mass star, these correspond to 3, 1, 0.3, and 0.1 Jupiter mass.}
\label{fig:numarm}
\end{figure*}

\section{WHICH CHARACTERISTICS OF OBSERVED SPIRAL ARMS CAN WE USE?}
\label{sec:characteristics}

The two main factors that determine the characteristics of spiral arms are the planet mass and the disk temperature.
If we can measure disk midplane temperature accurately, we may thus be able to use some characteristics of observed spiral arms to constrain the mass of unseen planet.
In order to examine which characteristics of spiral arms can be used, we carry out a suite of two-dimensional isothermal hydrodynamic simulations varying planet mass and disk temperature.

We solve the hydrodynamic equations for mass and momentum conservation in the two-dimensional polar coordinates $(r,\phi)$ using FARGO 3D \citep{benitez16}:
The simulation domain extends from $r_{\rm in} = 0.05~r_p$ to $r_{\rm out} = 5~r_p$ in radius and from 0 to $2\pi$ in azimuth.
We adopt 4096 logarithmically-spaced grid cells in the radial direction and 5580 uniformly-spaced grid cells in the azimuthal directions.

Our initial disk has a power-law surface density and temperature profile:  
$\Sigma_{\rm init}(r)   =  \Sigma_p \left({r/ r_p}\right)^{-1}$ and $T(r) = T_p \left( {r /r_p} \right)^{-1/2}$,
where $\Sigma_p$ and $T_p$ are the surface density and temperature at the location of the planet $r=r_p$.
We use planet masses of $M_p = 0.01, 0.1, 0.3, 1, 3$, and $10~M_{\rm th}$, where $M_{\rm th} \equiv c_s^3/\Omega G = M_* (h/r)_p^3$ is the so-called thermal mass \citep{lin93,goodman01}, at which planet mass its Hill radius is comparable to the disk scale height.
We choose $T_p$ such that the disk aspect ratio at the location of the planet is $(h/r)_p = 0.04, 0.05, 0.06, 0.07, 0.08, 0.1, 0.12$, and 0.15.
For $(h/r)_p$= 0.04 and 0.05, we also run calculations with $M_p = 30$ and $100~M_{\rm th}$.

In Figure \ref{fig:mosaic}, we display the perturbed surface density distributions with some selected parameters to provide an overview.
As expected, characteristics of spiral arms, such as the number of spiral arms, the pitch angle of spiral arms, and the arm-to-arm separation vary as a function of planet mass and disk temperature.

\begin{figure*}
\centering
\epsscale{1.05}
\plotone{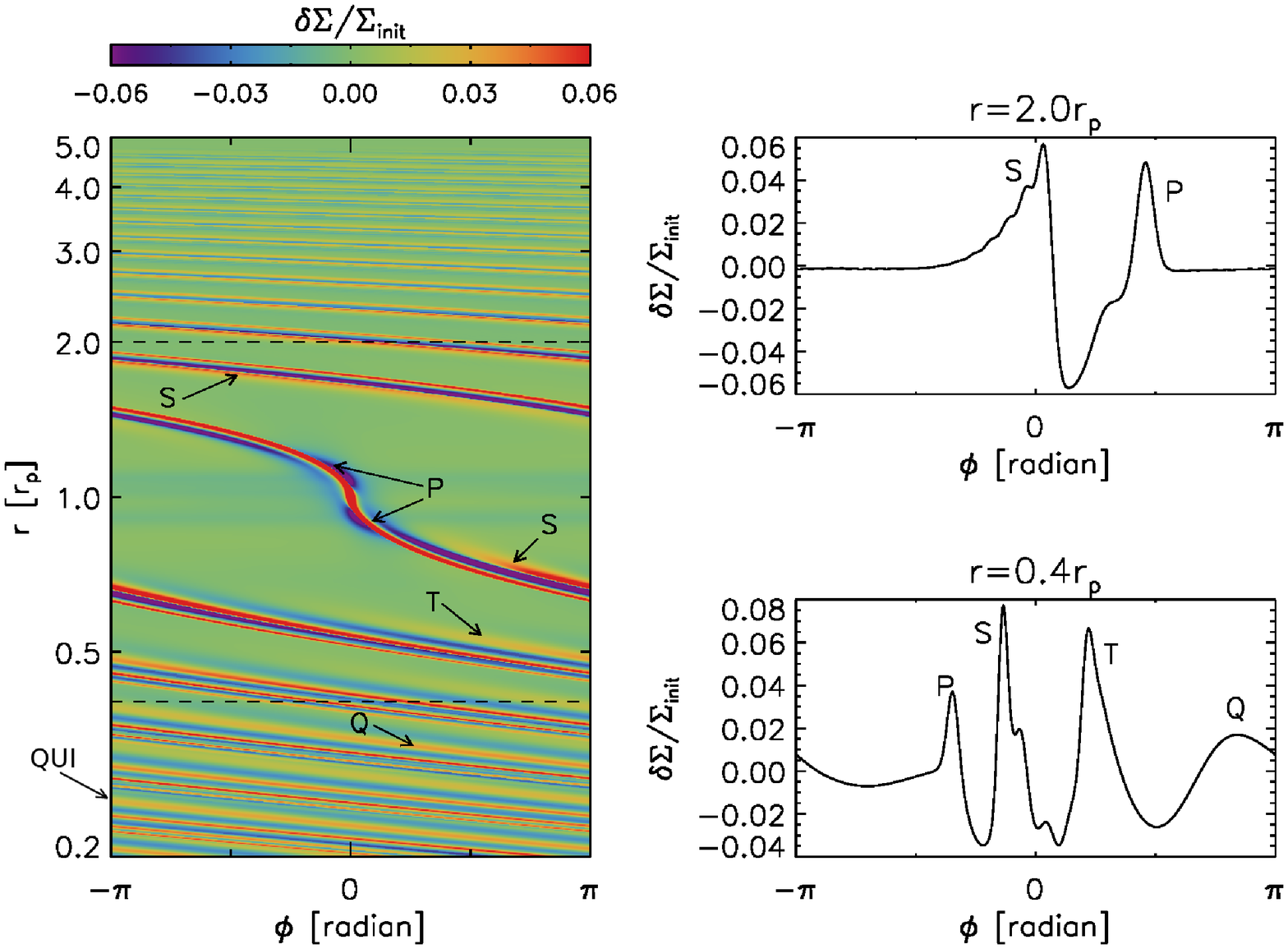}
\caption{(left) The two-dimensional perturbed density distribution $\delta\Sigma / \Sigma_{\rm init}$ with $(h/r)_p = 0.04$ and $M_p = 0.3~M_{\rm th}$ ($6.4~M_\earth$ assuming a solar-mass star). The horizontal dashed lines indicate where $r=0.4~r_p$ and $2.0~r_p$. (right) $\delta\Sigma / \Sigma_{\rm init}$ along azimuth at $r=0.4~r_p$ and $2.0~r_p$. Five spiral arms form in the inner disk and two spiral arms form in the outer disk. The primary, secondary, tertiary, quaternary, and quinary arms are indicated with `P', `S', `T', `Q', and `QUI', respectively.}
\label{fig:dens_hr004}
\end{figure*}

\subsection{Can We Use the Number of Observed Spiral Arms?}
\label{sec:numberofarms}

Additional spiral arms form when wave modes having different azimuthal wavenumbers become in phase as they propagate.
The propagation of wave modes depends on their azimuthal wavenumber but also the background disk temperature.
For a given azimuthal wavenumber, a wave mode in a colder disk is more tightly wound and narrower in width.
This raises the possibility that a larger number of additional arms can form.
In addition, as we showed in Paper I the mass of the planet also affects the number of spiral arms.

In Figure \ref{fig:numarm}, we present the number of spiral arms launched by planets in the inner ($0.2~r_p \leq r \leq 1~r_p$) and outer disk ($1~r_p \leq r \leq 5~r_p$) as a function of $(h/r)_p$ and $M_p$.
Planets launch two or more (up to five in our parameter space) spiral arms interior to their orbits.
At a given disk temperature, a smaller number of spiral arms form with a larger planet mass.
Also, in general, a smaller number of spiral arms form in a disk with a larger $(h/r)_p$.
As mentioned at the beginning of this section, this trend is expected because having a larger planet mass and/or a larger $(h/r)_p$ will make spiral arms (or wave modes) more opened, prevent more spiral arms forming.
Our parameter study shows that, for $0.04 \leq (h/r)_p \leq 0.15$, three or fewer spiral arms form interior to a planet's orbit  when $M_p/M_* \gtrsim 3\times10^{-4}$ and two-armed spirals form when $M_p/M_* \gtrsim 3\times10^{-3}$.
These masses correspond to a Saturn mass and 3 Jupiter masses, respectively, assuming a solar-mass central star.

Exterior to their orbits, we find that planets launch only one or two spiral arms.
The number of outer spiral arms appears to be more sensitive to the disk temperature than the planet mass.
Fewer spiral arms form in the outer disk than in the inner disk because propagation of wave modes become independent on the azimuthal wavenumber when $r \gg r_p$ (see Equation \ref{eqn:phase_m3}).
Therefore, additional spiral arms can form only when non-zero $n$th components become in phase within $r \lesssim {\rm a~few}~r_p$, which can occur in cold disks.
Our parameter study shows that two spiral arms form exterior to a planet's orbit only when $(h/r)_p \lesssim 0.06$.

In Figure \ref{fig:dens_hr004}, we present an example of which the largest number of spiral arms in our parameter study are launched: five in the inner disk and two in the outer disk. 
The parameters used in the example are $(h/r)_p=0.04$ and $M_p=0.3~M_{\rm th}$.
In addition to the primary arm directly attached to the planet, interior to the planet's orbit a secondary arm forms at $\sim0.8~r_p$, a tertiary arm forms at $\sim0.6~r_p$, a quaternary arm forms at $\sim0.4~r_p$, and a quinary arm forms at $\sim0.3~r_p$.
Exterior to the planet's orbit, a secondary arm forms at $\sim1.5~r_p$ and no more spiral arms form beyond this radius.
With a small planet mass of $0.3~M_{\rm th}$ (=6.4 Earth mass assuming a solar mass star), the level of perturbation driven by the spiral arms is low ($\delta\Sigma/\Sigma_{\rm init} \lesssim 0.1$).
These spiral arms are therefore unlikely to be detectable in near-IR scattered light images  \citep{dongfung17}.
While weak, however, these spiral arms may still be able to generate observable signatures preferentially in low-viscosity disks by opening gaps as they shock the disk gas, provided that sufficient time is allowed (\citealt{bae17}; see also Section \ref{sec:gaps} of the present paper).

\begin{figure*}
\centering
\epsscale{1.1}
\plotone{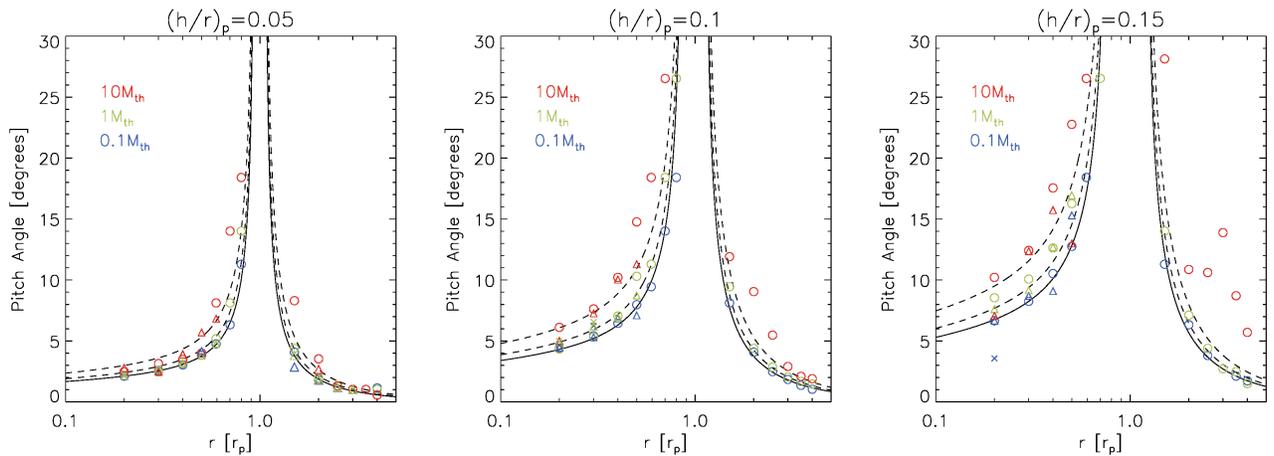}
\caption{Pitch angle of the (circle) primary, (triangle) secondary, and (cross) tertiary arm for (blue)  $0.1~M_{\rm th}$, (green) $1~M_{\rm th}$, and (red) $10~M_{\rm th}$ planets. Each panel presents results with (left) $(h/r)_p=0.05$, (middle) $(h/r)_p=0.1$, and (right) $(h/r)_p=0.15$. The black solid curves present the pitch angle calculated with the dominating azimuth mode predicted by linear theory $m=(1/2)(h/r)_p^{-1}$ in the spiral arm phase equation (Equation \ref{eqn:phase_m3}). The lower and upper dashed curves in each panel show the predicted pitch angles of spiral arms by linear theory with $30~\%$ and $100~\%$ higher disk temperature.}
\label{fig:pitch_tot}
\end{figure*}

\subsection{Can We Use the Pitch Angle of Spiral Arms?}
\label{sec:pitch_angle}

In Figure \ref{fig:pitch_tot}, we present measured pitch angles of spiral arms from our simulations.
The pitch angles $\beta$ are measured using $\tan \beta = -{\rm d}r/r{\rm d} \phi$ as we follow each spiral arm in radius.
In general, for any given disk temperature, larger planet masses result in larger pitch angles.
So in theory we can use the pitch angle to constrain the mass of unseen planet. 
In practice, however, there are difficulties.
First of all, we note that highly accurate temperature measurements are required.
As shown in Figure \ref{fig:pitch_tot} even a $30~\%$ uncertainty in temperature can change the mass estimates by an order of magnitude. 
Given the difficulty of measuring the actual midplane gas temperature, arising for instance from relating mm continuum emissions to gas temperatures or from using molecular line ratios which tend not to probe exactly the same disk regions in height, making an one-to-one correlation between the pitch angle of a spiral arm to the planet mass seems challenging, at least for now.
Even when the disk midplane temperature is accurately measured, an additional uncertainty in the mass constraints comes from the fact that we do not know the location of the planet and thus the exact $(h/r)_p$ value.
Many of the spiral arms observed so far have fluctuating pitch angles \citep[e.g.,][]{reggiani17}, another challenge to use the pitch angle to constrain planet mass.

On the other hand, as shown in Figure \ref{fig:pitch_tot} the pitch angles of both primary and additional spiral arms decrease away from the planet.
This is expected from linear wave theory and numerical simulations show the trend as well.
It is a particularly important characteristic because, if we measure the pitch angle of a spiral arm over a large enough radial range, the increasing or decreasing trend of the pitch angle can be used to distinguish whether the spiral arm is excited by a planet inside or outside of the spirals (see Section \ref{sec:mwc758}).

\subsection{Can We Use the Arm-to-Arm Separation?}
\label{sec:arm_separation}

Previous numerical simulations showed that the separation between the primary and secondary arm increases with the planet mass \citep{zhu15,fung15,lee16}.
Interestingly, a parameter study presented in \citet{fung15} showed that the arm-to-arm separation is independent of the disk temperature profile.
It is thus suggested that the arm-to-arm separation can be broadly used to infer the mass of unseen planet.

\begin{figure}
\centering
\epsscale{1.12}
\plotone{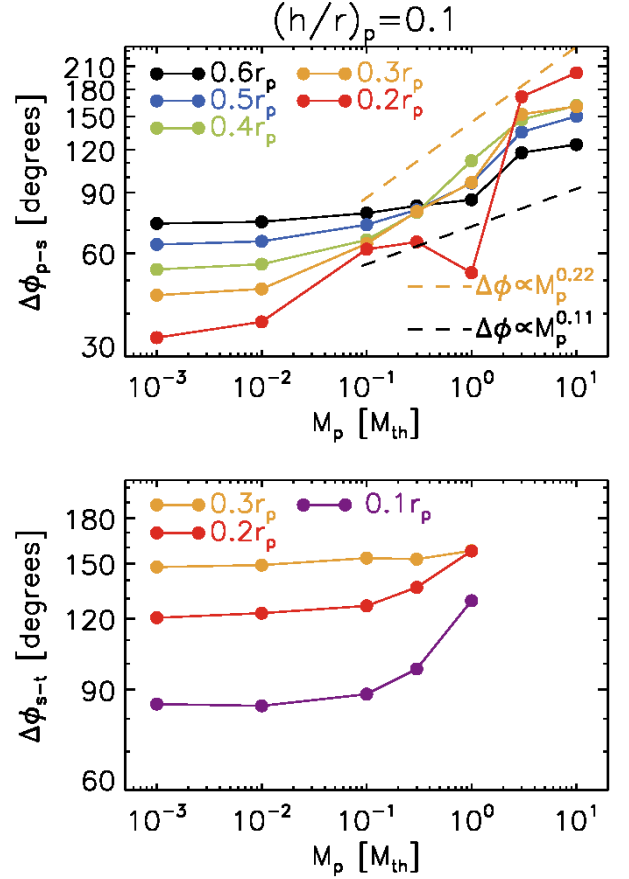}
\caption{The angular separation between (top) primary and secondary arms $\Delta\phi_{p-s}$ and (bottom) secondary and tertiary arms $\Delta\phi_{s-t}$, as a function of the planet mass. Disk aspect ratio of $(h/r)_p=0.1$ is used. Note that the arm-to-arm separation increases as a function of planet mass in general. The best fits to the data points with $M_p  \geq 0.1~M_{\rm th}$ ($M_p/M_* \geq 10^{-4}$) at $r=0.3$, 0.4, 0.5, and $0.6~r_p$ in the top panel have slopes of 0.22, 0.21, 0.17, and 0.11, respectively. For 0.3 and $1~M_{\rm th}$, the measured $\Delta\phi_{p-s}$ values are off from the general trend at $r=0.2~r_p$ because of the interference between the spiral arms (see Section 4 of Paper I).}
\label{fig:pdiff}
\end{figure}

We compute arm-to-arm separations from our simulations.
Specifically, we measure the angular separation between spiral arms at each radius.
We then average it around $r=0.2$, 0.3, 0.4, 0.5, and $0.6~r_p$ for the primary-to-secondary separation and around $r=0.1$, 0.2, and $0.3~r_p$ for the secondary-to-tertiary separation, with a radial bin of $\Delta r =0.1r_p$.
Focusing on $(h/r)_p=0.1$ models first, we present the primary-to-secondary separation $\Delta\phi_{p-s}$ and the secondary-to-tertiary separation $\Delta\phi_{s-t}$ as a function of the planet mass in Figure \ref{fig:pdiff}.
For both $\Delta\phi_{p-s}$ and $\Delta\phi_{s-t}$, there is a general trend that the arm-to-arm separation increases with the planet mass, consistent with previous studies \citep{zhu15,fung15}.
For $\Delta\phi_{p-s}$, it converges to $\sim 180^\circ$ beyond $M_p \gtrsim 3~M_{\rm th}$ for which planet mass only two spiral arms form.
For the planet masses in a range of $M_p \geq 0.1~M_{\rm th}$ ($M_p/M_* \geq 10^{-4}$), the primary-to-secondary arm separation at $0.2~r_p \leq r \leq 0.6~r_p$ is best fitted with $\Delta\phi_{p-s} = 101^\circ (M_p/M_{\rm th})^{0.20}$.
However, we note that the arm-to-arm separation increases more steeply with the planet mass at smaller radii. 
The best-fit slopes for individual radial bins increase from 0.11 at $0.6~r_p$ to 0.22 at $0.3~r_p$.
Similarly, the slope in $\Delta\phi_{s-t}$ also increases more steeply at smaller radii.
It is therefore helpful to know an approximate planet location for more accurate mass estimates with arm-to-arm separation.

\begin{figure*}
\centering
\epsscale{1.12}
\plotone{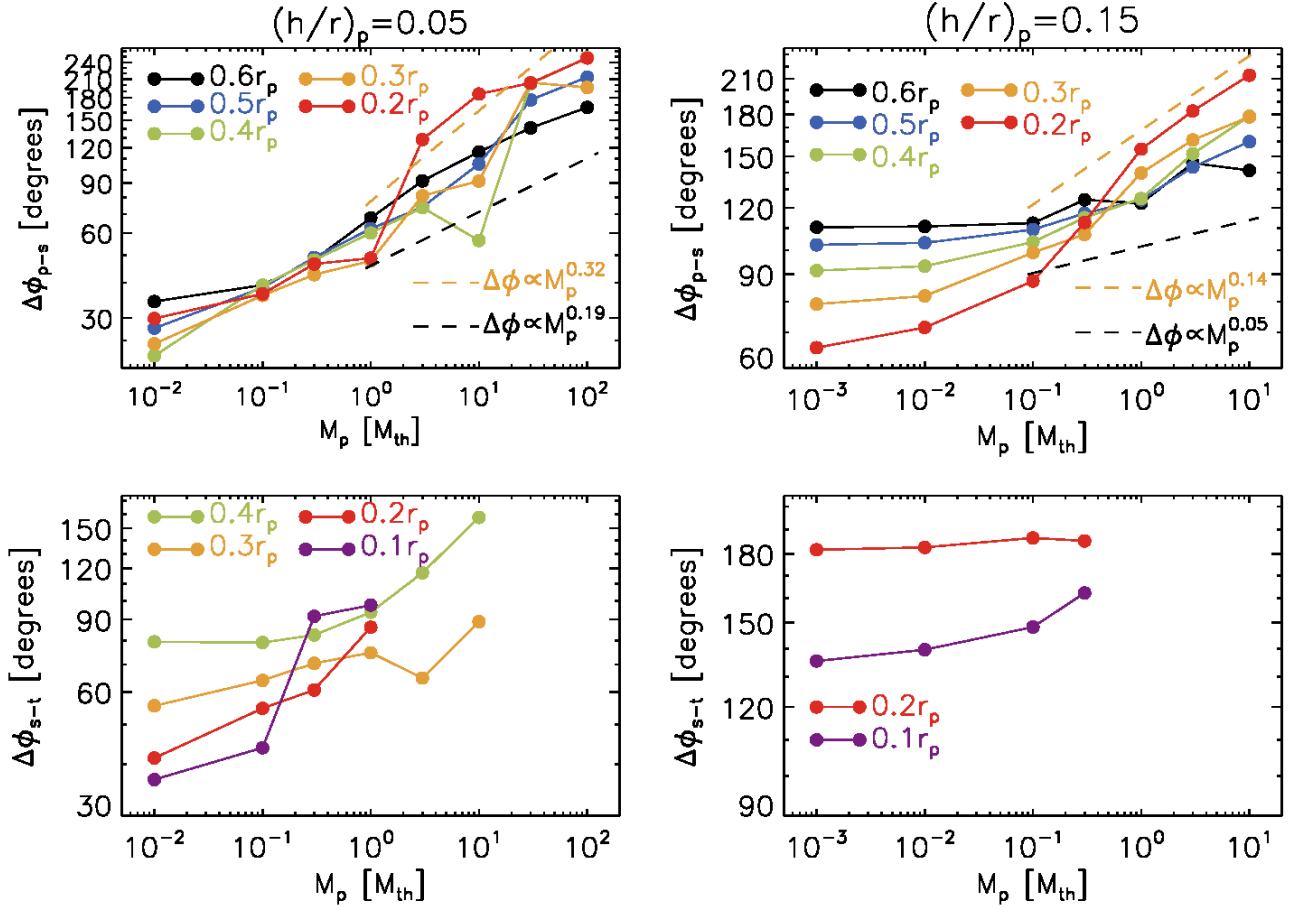}
\caption{Same as Figure \ref{fig:pdiff} but with (left panels) $(h/r)_p=0.05$ and (right panels) $(h/r)_p=0.15$. For $(h/r)_p=0.05$, spiral arms driven by a $M_p=10^{-3}~M_{\rm th}$ ($q=6.4\times10^{-8}$) planet are too weak to precisely measure the arm-to-arm separation.  For $M_p=3$ and $10~M_{\rm th}$ with $(h/r)_p=0.05$, spiral arms merge and only two arms are left at $r \lesssim 0.2~r_p$.}
\label{fig:pdiff2}
\end{figure*}

There are a few other things worth pointing out from Figure \ref{fig:pdiff}.
First, compared at a given radius, $\Delta\phi_{p-s}$ is larger than $\Delta\phi_{s-t}$ regardless of the planet mass.
Second, both $\Delta\phi_{p-s}$ and $\Delta\phi_{s-t}$ flatten out for sufficiently low mass planets, suggesting that spiral arms cannot be infinitely close to each other in the linear regime.
Last, for a given planet mass, $\Delta\phi_{p-s}$ increases as a function of radius for low-mass planets (i.e., $M_p < 0.3~M_{\rm th}$) while the separation decreases as a function of radius for high-mass planets (i.e., $M_p > 0.3~M_{\rm th}$).

In Figure \ref{fig:pdiff2}, we present $\Delta\phi_{p-s}$ and $\Delta\phi_{s-t}$ measured in simulations with $(h/r)_p=0.05$ and $(h/r)_p=0.15$.
Comparing Figure \ref{fig:pdiff2} with Figure \ref{fig:pdiff}, we note that the general trends seen in the fiducial case with $(h/r)_p=0.1$ are commonly seen with $(h/r)_p=0.05$ and 0.15.
Most importantly, the arm-to-arm separation increases more steeply with the planet mass at smaller radii with other $(h/r)_p$ values.
It is worth noting that this trend is more clearly seen with larger $(h/r)_p$ values because the interference and/or merging between spiral arms tends to occur more frequently with smaller $(h/r)_p$ values.

\begin{figure}
\centering
\epsscale{1.15}
\plotone{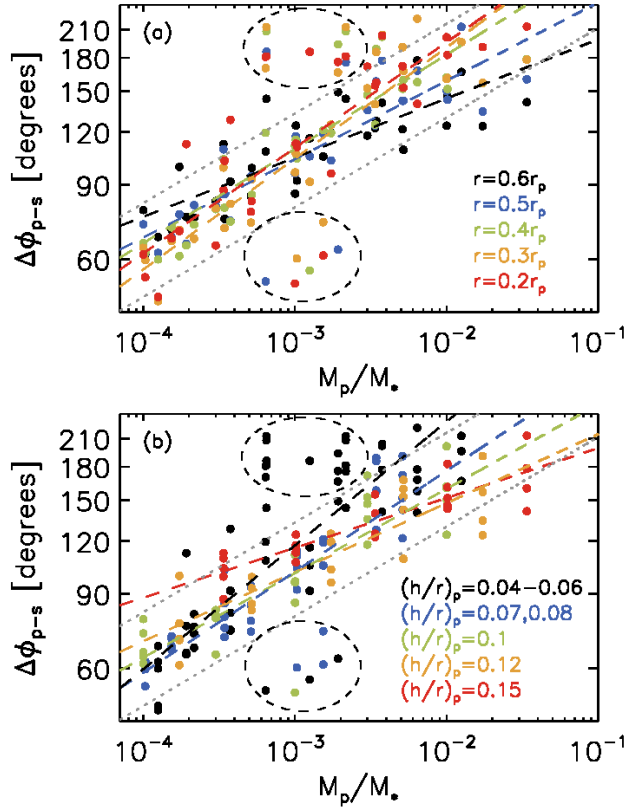}
\caption{(a) The angular separation between the primary and secondary arm $\Delta\phi_{p-s}$ measured at different radial bins from all models. The data points are color coded with the radii the separations are measured. (b) Same as (a) but data points are color coded with $(h/r)_p$ values. The dashed lines in the panels present the best-fit to the data points with the same color. The gray dotted lines present the best-fit to the all data points: $\Delta\phi_{p-s} \propto (M_p/M_*)^{0.21}$. See Table \ref{tab:fits} for the best fits. The data points in the upper dashed ellipse in each panel experience merging between spiral arms. The data points in the lower dashed ellipse in each panel experience interference between spiral arms such that the spiral arms are more closely located in azimuth than the general trend.}
\label{fig:pdiff_tot}
\end{figure}

In Figure \ref{fig:pdiff_tot} (a), we present $\Delta\phi_{p-s}$ computed in all our models differing $M_p$ and $(h/r)_p$.
Data points are color-coded to represent $\Delta\phi_{p-s}$ at different radial bins.
Note that the $x$ axis of the plot is now planet-to-star mass ratio $M_p/M_*$, different from Figure \ref{fig:pdiff} and \ref{fig:pdiff2}.
We fit the data points having $M_p/M_* \geq 10^{-4}$, chosen similarly to \citet{fung15} for a comparison.
The entire data points are best fitted with $\Delta\phi_{p-s} = 106^\circ (q/0.001)^{0.21}$ where $q\equiv M_p/M_*$, which is in an excellent agreement with \citet{fung15}.
However, as in Figure \ref{fig:pdiff} and \ref{fig:pdiff2}, we find there is a general trend that the arm-to-arm separations increase more steeply at smaller radii.
In Figure \ref{fig:pdiff_tot} (b), data points are now color-coded to represent different $(h/r)_p$ in models.
Again, we do find a trend that the arm-to-arm separation increases more steeply with smaller $(h/r)_p$ values.
In Table \ref{tab:fits}, we provide the best fits to $\Delta\phi_{p-s} - M_p/M_*$ relation.
A fit given for a certain radius uses the data points with all $(h/r)_p$ values, whereas a fit given for a certain $(h/r)_p$ value uses the data points at all different radii. 
In the appendix, we provide best fits for each $(h/r)_p$ values and radial bins.

To summarize, we confirm that the arm-to-arm separation increases with the planet mass, in agreement with previous studies \citep{zhu15,fung15,lee16}.
Our best-fit of primary-to-secondary separation is $\Delta\phi = 106^\circ (q/0.001)^{0.21}$.
However, we find that, in the inner disk, the arm-to-arm separation increases more steeply with the planet mass at smaller radii and with smaller $(h/r)_p$ values.

\begin{deluxetable}{cl}
\tablecolumns{2}
\tabletypesize{\small}
\tablecaption{Best Fits to $\Delta\phi_{p-s} - M_p/M_*$ relation \label{tab:results}}
\tablewidth{0pt}
\tablehead{
\colhead{Data} & 
\colhead{Best Fits} 
}
\startdata
entire data points & $\Delta\phi_{p-s} = 106^\circ (q/0.001)^{0.21}$ \\
\hline
$r = 0.2~r_p$ & $\Delta\phi_{p-s} = 110^\circ (q/0.001)^{0.25}$ \\
$r = 0.3~r_p$ & $\Delta\phi_{p-s} = 104^\circ (q/0.001)^{0.26}$ \\
$r = 0.4~r_p$ & $\Delta\phi_{p-s} = 109^\circ (q/0.001)^{0.22}$ \\
$r = 0.5~r_p$ & $\Delta\phi_{p-s} = 104^\circ (q/0.001)^{0.19}$ \\
$r = 0.6~r_p$ & $\Delta\phi_{p-s} = 104^\circ (q/0.001)^{0.14}$ \\
\hline
$(h/r)_p=0.04, 0.05, 0.06$ & $\Delta\phi_{p-s} = 117^\circ (q/0.001)^{0.29}$ \\
$(h/r)_p=0.07, 0.08$ & $\Delta\phi_{p-s} = 102^\circ (q/0.001)^{0.24}$ \\
$(h/r)_p=0.1$ & $\Delta\phi_{p-s} = 101^\circ (q/0.001)^{0.20}$ \\
$(h/r)_p=0.12$ & $\Delta\phi_{p-s} = 101^\circ (q/0.001)^{0.16}$ \\
$(h/r)_p=0.15$ & $\Delta\phi_{p-s} = 119^\circ (q/0.001)^{0.10}$ 
\enddata
\end{deluxetable}

\subsection{Can We Use the Relative Brightness/Intensity of Spiral Arms?}

First of all, secondary and tertiary arms can create stronger density perturbations/contrast than the primary arm at a given radius (e.g., Figure \ref{fig:dens_hr004}; see also \citealt{fung15}). 
In addition, the brightness/intensity of spiral arms can significantly differ from their intrinsic brightness/intensity because of interaction with background disk structures (e.g., a vortex; \citealt{bae16a}) and/or distortion via the spiral wave instability \citep{bae16b}.
In case of near-IR scattered light observations, shadows cast by structures located closer to the central star may also affect the brightness of spiral arms located far away \citep{stolker16}.
We thus caution that observed intensity or brightness of spiral arms is not an ideal characteristic to determine whether a certain spiral arm is a primary or a secondary/tertiary but also to provide planet mass constraints.
Variations in the brightness of spiral arms seen in multi-epoch scattered light observations \citep[e.g.,][]{stolker16,stolker17} also support this conclusion.

\section{Application to the Observed Spiral Arms}
\label{sec:application}

\subsection{MWC~758}
\label{sec:mwc758}

MWC~758 is a $3.5\pm2$~Myr-old Herbig Ae star \citep{meeus12} at a distance of $151^{+9}_{-8}$~pc \citep{gaia16}.
Two spiral arms are detected in near-IR scattered light observations using the Subaru Telescope High Contrast Instrument with Adaptive Optics in K$_s$- and H-band \citep{grady13} and the VLT with Spectro-Polarimetric High-contrast Exoplanet Research in Y-band \citep{benisty15}.
Recently, using L'-band observations with NIRC2 at the Keck II telescope, \citet{reggiani17} reported three spiral arms (two previously reported + one new) spanning in between $\sim$0."25 and $\sim$0."6 ($\sim40-90$~au) from the central star.
In addition, a point-like source is detected at a separation of 0."11 ($\sim$20~au) from the central star \citep{reggiani17}, making the disk a very interesting object to test planet-disk interaction theories.

To examine a potential planetary origin of the spiral arms in the MWC~758 disk, we first assume an external companion at 100 au and $h/r=0.12$ at the radius, similarly to the model in \citet{dong15} but rescaled with the distance based on \citet{gaia16}.
Assuming that the external companion is responsible for all three spiral arms, our parameter study suggests that the unseen planet should have a mass less than 3 thermal mass; a larger-mass planet would excite only two spiral arms.
Adopting a 1.5 solar mass central star \citep{isella10,reggiani17} gives an upper limit of 7.8 Jupiter mass.
An interesting feature regarding the arm-to-arm separation is that the separation between the arm S1 and S3 decreases with radius (see Figure 7 of \citealt{reggiani17}).
Although a radially decreasing arm-to-arm separation is not completely unexpected as Figure \ref{fig:pdiff} shows (see $M_p=3$ and $10~M_{\rm th}$ for example), such a small arm-to-arm separation of $\sim45^\circ$ as well as a rapid drop over a short radial distance is not what is typically seen our simulations -- assuming again a companion at 100~au as above, the arm-to-arm separation decreases from $140^\circ$ at $\sim0.25~r_p$ to $45^\circ$ at $\sim0.63~r_p$.
One possibility to explain these features might be interference between the arms (Paper I), but our understanding of the phenomenon is too incomplete to arrive at a conclusion.

We note that the point-like source detected in \citet{reggiani17} is unlikely to be the perturber exciting three spiral arms exterior to its orbit based on our parameter study.
However, it is possible that the point-like source excites one of the three spiral arms and a yet unseen external companion excites the other two. 
In fact, we find some supporting features that S3 could be an outer spiral of a planet interior to S3.
First, the decreasing pitch angle of S3 as a function of radius (Figure 5 of \citealt{reggiani17}) supports the idea that S3 is an outer spiral arm launched by an internal perturber (see Section \ref{sec:pitch_angle}).
This also helps resolve the arm-to-arm separation problems discussed in the previous paragraph.
In addition, it is known from three-dimensional simulations that inner spiral arms produce significant vertical motion while the outer spiral arms induce little vertical motion \citep{zhu15}.
If S3 is an outer spiral arm it is likely that the pressure scale height along the arm does not increase significantly.
This can explain the non-detection of S3 at a shorter wavelength in Y-band ($1.04~\mu$m; \citealt{benisty15}) which trace upper layers of the disk than L'-band.
This scenario is consistent with a suggestion made in \citet{reggiani17} and also supports the suggestion by \citet{juhasz15} that observed spiral arms might be the results of pressure scale height changes.
While no apparent physical connection between the point-like source and S3 is found from the scattered light image, it is possibly because the point-like emission traces a protoplanet located near the midplane whereas S3 traces the surface.
This explanation is consistent with the inclination and PA of the disk.
In the case an unseen external planet excites two of the three spiral arms, the external companion has to be more massive than a thermal mass (see Section \ref{sec:numberofarms}).
We are thus left with a fairly narrow companion mass range of $2.6-5$ Jupiter mass: the lower limit from our parameter study and the upper limit from \citet{reggiani17}.

When the three spiral arms do not share the same origin, distinguishable orbital motions among the spirals are expected in monitoring observations.
If spiral arm S3 is excited by the point-like source at 20~au, the spiral pattern will rotate $4.9^\circ$ per year in the deprojected plane. 
Assuming an external planet at $\gtrsim100$~au is responsible for the other two spiral arms, the two spirals will rotate $\lesssim0.4^\circ$ per year in the deprojected plane.
Therefore, observations in 2018 may reveal more than $>10^\circ$ of difference in the orbital motions between spiral arm S3 and the other two compared with the first epoch observation of \citet{reggiani17} taken in 2015 October.

\subsection{Elias 2-27}

Elias 2-27 is a young ($\sim1$~Myr; \citealt{luhman99}), low-mass (0.6~solar-mass; \citealt{andrews09}) star in the $\rho$-Ophiuchus star-forming region.
This object is particularly interesting because spiral arms are detected at mm wavelengths with ALMA \citep{perez16}.
The emission at 1.3~mm is optically thin, so the continuum emission traces down to the midplane of the disk \citep{perez16}, in contrast to near-IR scattered light imaging which trace spiral arms near the disk surface.

\begin{figure}
\centering
\epsscale{1.15}
\plotone{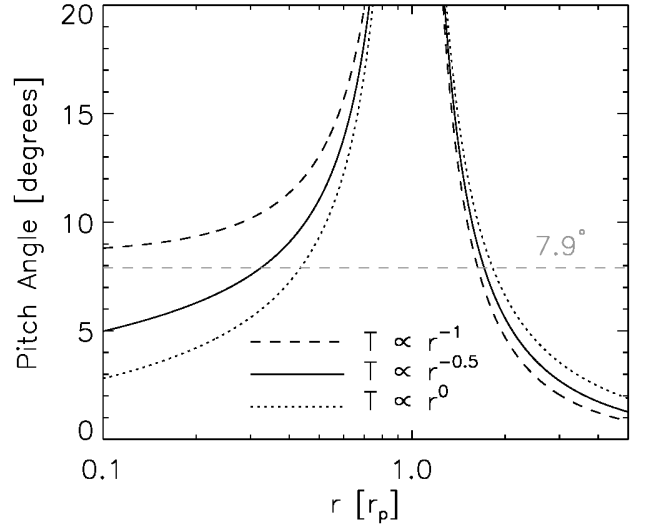}
\caption{Pitch angle of a spiral arm predicted by the linear wave theory assuming $(h/r)_p=0.15$. The three curves assume different temperature profiles: (dashed) $T\propto r^{-1}$, (solid) $T \propto r^{-0.5}$, and (dotted) $T \propto r^0$. Over-plotted with a gray dashed line indicates a pitch angle of $7.9^\circ$, the best-fit value derived for the two-armed spirals in the Elias~2--27 disk \citep{perez16}.}
\label{fig:pitch_elias}
\end{figure}

Both spiral arms in the Elias 2-27 disk appear to be well fit with two symmetric logarithmic spirals, i.e., a constant pitch angle of $7.9^\circ \pm 0.4^\circ$ \citep{perez16}.
This is an interesting characteristic because a constant pitch angle is unlikely for planet-driven spiral arms (Figure \ref{fig:pitch_tot}; see also \citealt{zhu15}).
One possibility is that the disk has a steeper temperature gradient than the constraints made in \citet{perez16}.
In Figure \ref{fig:pitch_elias} we plot the predicted pitch angles as a function of radius based on the linear wave theory, assuming $(h/r)_p=0.15$.
As shown, the steeper the disk temperature gradient is, the more slowly the pitch angle varies in the inner disk.
If it is the temperature gradient alone though, the planet probably has to be located at very large distance ($\gtrsim 500$~au) to help its inner spiral arms have a constant pitch angle.
More accurate temperature measurements in the future could help test whether this is the case.
If the two spiral arms are driven by a planet exterior to the spirals, our parameter study suggests that the planet has to be more massive than about a Jupiter mass assuming the midplane temperature constrained by \citet{perez16}.
This planetary mass is consistent with constraints made by previous observations and numerical simulations (see \citealt{meru17} and references therein).
The disk has $h/r \geq 0.06$ beyond $\sim10$~au, so it is very unlikely that an internal planet excites the two spiral arms.

Alternatively, it might be that a constant pitch angle is a generic characteristic of GI-driven spiral arms.
In fact, the GI-driven spiral arms seen in numerical simulations of \citet{bae14} have a constant pitch angle over a broad range of radius, in particular inward of the gravitationally unstable region\footnote{The GI-driven two-armed spiral arms in Figure 14 of \citet{bae14} are well fitted with logarithmic spirals having a pitch angle of $\sim7^\circ$ over $r=0.5-2.5$~au.}.
If the shear rate of the disk rotation is what determines the pitch angle of GI-driven spiral arms, the potential universality of a constant pitch angle among GI-driven spiral arms might be explained.
\citet{perez16} calculated the Toomre $Q$ parameter based on the dust continuum emission and suggested that the disk can be gravitationally unstable at the radii where the spiral arms are detected, when the dust opacity is reduced by a factor of $\gtrsim 4$ than what is typically assumed in literature. 
More accurate observational constraints on the gas surface density as well as improvements in our understanding of GI-driven spiral arms are desired to better understand the origin of the spiral arms in the Elias 2--27 disk.

\section{Spiral Arms as an Origin of Concentric Rings and Gaps}
\label{sec:gaps}

It has been suggested that planet-driven spiral arms can create multiple concentric rings and gaps as they shock disk gas at different radial locations in a disk \citep{bae17}.
While we have considered the appearance of planet-driven spiral arms in previous sections, in certain circumstances rings and gaps can have more dominant observable signature than spiral arms themselves.
In this section, we discuss when we expect to observe rings and gaps rather than spiral arms, and vice versa.
We then discuss the generation of multiple rings and gaps in the solar nebula by a proto-Jupiter and its potential implications.

\subsection{Spiral arms vs. rings and gaps}

In order for spiral arms to be observable, they have to (1) have a large enough pitch angle to overcome a finite spatial resolution (see e.g., Figure 4 of \citealt{kanagawa15}); and (2) produce a sufficient level of perturbations \citep{dongfung17}.
For given disk property the two conditions can be more readily met with a large planet mass, since spiral arms from a more massive planet have a larger pitch angle (Figure \ref{fig:pitch_tot}) and produce a larger level of perturbations (Figure \ref{fig:mosaic}).
While \citet{dongfung17} noted that assessing the detectable planet mass limit generally requires a case-by-case study, their empirical scaling relation between arm-to-disk contrast and planet mass suggests that multi-Jupiter-mass planets are required to produce the arm-to-disk contrast of the observed spiral arms (e.g., MWC~758, SAO~206462).
In agreement with the arm-to-disk contrast argument, hydrodynamic modeling of observed spiral arms suggests that order of 10 Jupiter-mass planets are required to reproduce the morphology of nearly axisymmetric $m=2$ spiral arms: $9~M_{\rm Jup}$ for MWC~758 \citep{dong15}; $10~M_{\rm Jup}$ for SAO~206462 \citep{bae16a}; and Elias~2-27 \citep{meru17}.

The disk viscosity can also affect the observability of spiral arms versus rings and gaps.
The morphology of planet-driven spiral arms and the arm-to-disk contrast are shown to be not very sensitive to disk viscosity \citep{dongfung17}.
On the other hand, spiral arms are not able to open gaps when the mass transport by the background disk turbulence exceeds that by spiral shocks.
While the parameter space has not been fully explored yet, the results in \citet{bae17} suggest that a viscosity of $\alpha \lesssim 10^{-3}$ is required for a Jupiter-mass planet to create multiple rings and gaps.
Thus, rings and gaps can be observed preferentially in disks with low viscosity.

The method through which substructures are probed is also important. 
Probing disk surface layers using near-infrared scattered light or optically thick line emissions will offer a higher chance of detecting spiral arms, because significantly larger perturbations are expected from spiral arms at the disk surface than at the disk midplane \citep{zhu15}.
On the contrary, (sub-)mm dust continuum traces deep in the disk near the midplane, where the perturbations driven by spiral arms are intrinsically smaller.
In addition, as millimeter-sized particles experience more significant aerodynamic drag than $\mu$m-sized particles, they can be more efficiently collected in pressure bumps rather than collected at the spiral arm front.

Lastly, timescale also matters. 
Spiral arms launch and reach a quasi-steady state within a sound-crossing time, which is typically a few planetary orbital time.
On the other hand, rings and gaps require much longer time to fully develop \citep[$\gtrsim$ 100 orbital times,][]{bae17}.

\begin{figure*}
\centering
\epsscale{1.15}
\plotone{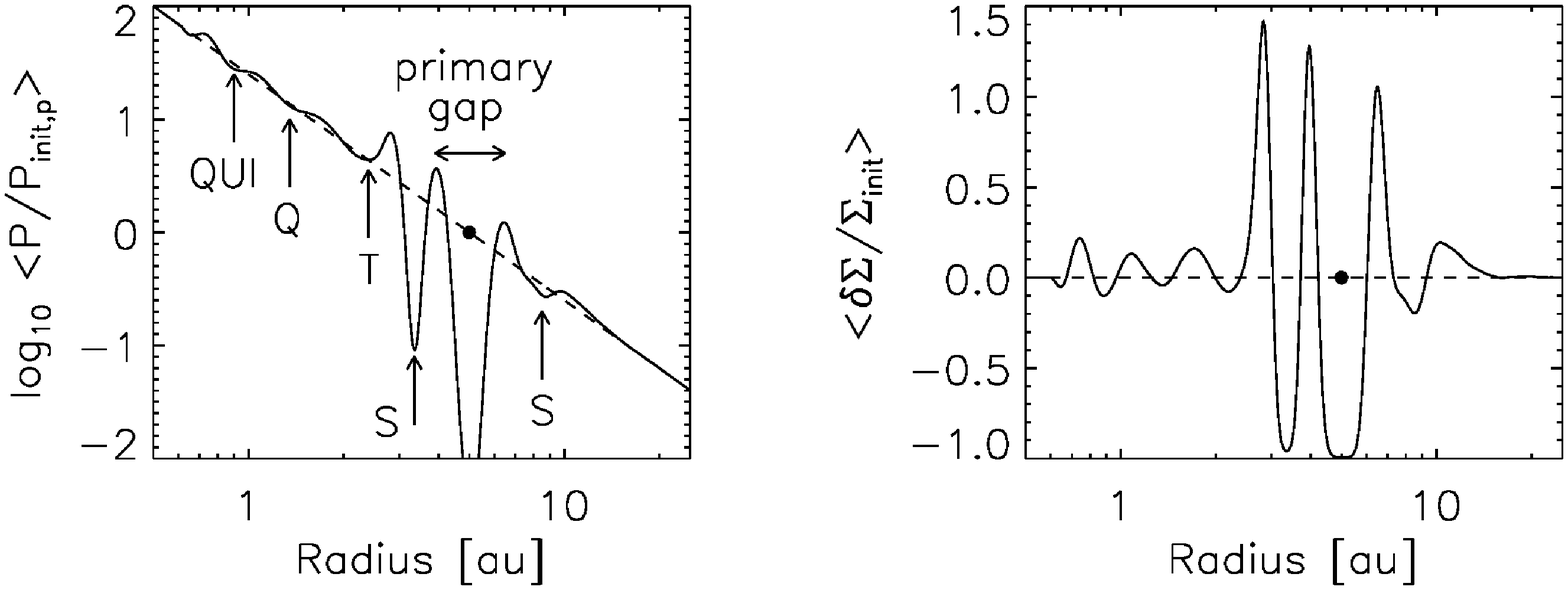}
\caption{(Left) The azimuthally averaged radial gas pressure profile in a logarithmic scale, after 200~kyr of evolution. It is normalized by the initial gas pressure at the core's location $P_{\rm init, p}$. The $20~M_\earth$ core opens a primary gap around its orbit. In addition to the primary gap each spiral arm opens a gap as it steepens into a shock, indicated with `S', `T', `Q', and `QUI'. The innermost gap-like density drop at $r\sim0.65$~au may be due to another spiral arm launching at the radius, but since it is located too close to the inner damping zone ($r=0.5-0.6$~au) we do not classify it as a physical gap. As such, we do not classify the pressure bump at $r\sim0.75$~au as a physical bump. (Right) The radial distribution of the azimuthally averaged perturbed density. Note that the core generates at least four pressure bumps ($r=2.8$, 4.0, 6.5, and 10~au) that are strong enough to trap solid particles. The core is presented with a filled black circle at 5~au.}
\label{fig:jupiter}
\end{figure*}

\subsection{Potential Implications in the Solar Nebula}
\label{sec:solar_nebula}

Meteorites contain calcium-aluminum-rich inclusions (CAIs), millimeter- to centimeter-sized particles believed to be formed very early in the solar nebula \citep[e.g.,][]{connelly12}.
Interestingly, meteorite parent body accretion is known to have continued for 2-4~Myr after the formation of CAIs \citep{connelly12,kita12,budde16b}.
Millimeter- to centimeter-sized particles would experience very rapid inward drift through aerodynamic drag.
It has been suggested that one way to store CAIs in the solar nebula for millions of years is if they were trapped in local pressure maxima until they accumulate in to meteorite parent bodies \citep{desch17}.
Moreover, there are different types of meteorites each of which has distinct chemical compositions and CAI sizes and abundances, suggesting that there might have been multiple spatially separated reservoirs of solid particles in the solar nebula \citep{warren11,budde16a,kruijer17}.

Here we examine the possibility that Jupiter's growing core could create multiple pressure maxima in the solar nebula, which could have acted as solid particle reservoirs.
In \citet{bae17}, we showed that spiral arms driven by a planet (or a planetary core) can create multiple gaps in the disk.
Pressure maxima (i.e., rings) can develop between those gaps, potentially trapping solid particles.

Following the conclusion of \citet{kruijer17} that Jupiter's core had to grow quickly to $20~M_\earth$ within 1~Myr, making it to be the oldest planet of the solar system, we carry out a numerical simulation with a $20~M_\earth$ core located at 5~au.
We adopt a minimum mass solar nebula surface density profile  \citep{weidenschilling77,hayashi81} and a passively heated, stellar irradiation-dominated temperature profile: $T(r) = 180~{\rm K}(r/1{\rm au})^{-1/2}$.
With this disk model $(h/r)_p=0.04$ at 5~au and the $20~M_\earth$ Jupiter's core is $94~\%$ of a thermal mass.
We use the inner and outer boundaries of 0.5 and 25~au and adopt 1024 logarithmically-spaced grid cells in radius and 1640 uniformly-spaced grid cells in azimuth.

The radial gas pressure distribution after 200~kyr ($\sim$18000 orbits) of evolution is shown in Figure \ref{fig:jupiter}.
The core excites five spiral arms interior to its orbit and two spiral arms exterior to its orbit.
Each spiral arm opens a gap through shock dissipation and pressure maxima form in between the gaps.
In total, the core creates four pressure bumps interior to its orbit and two pressure bumps exterior to its orbit.

To test whether these pressure bumps are strong enough to trap solid particles, we follow the analytic approach presented in \citet{zhu12}.
To briefly summarize, in order for a particle to be trapped in a pressure bump the drift velocity toward the pressure maximum has to be greater than two velocities: (1) the diffusion velocity by gas turbulence; and (2) the enhanced radial gas velocity within the gap caused by the deficit of gas.
The second requirement assumes a constant mass accretion rate across the ring and gap.
One assumption has to be made to use this analytic approach is that the shape of additional gaps are similar to the main gap within which the planet orbits, as described by Equation (24) in \citet{zhu12}.

Using $T_{s,0} = \rho_p s \pi / (2 \Sigma_{g,0})$, Equation (27) of \citet{zhu12} can be rearranged as 
\begin{equation}
\alpha_{\rm diff}  = {{\rho_p s \pi} \over {2 \Sigma_g}} \left[ \ln \left( \gamma_0 \over \gamma \right) \right]^{-1},
\end{equation}
where $\rho_p$ is the solid particle density, $s$ is the radius of the solid particle, $\Sigma_g$ is the gas surface density, $\gamma_0/\gamma$ is the dust depletion factor.
Now, let us consider a mm-sized particle placed right outside of the inner tertiary gap at $\sim 2.3$~au to examine whether it will be dragged into the inner disk or dragged outward to be trapped at the inner secondary pressure bump at $\sim2.8$~au, as an example.  
Considering a spherical solid particle with a density of $\rho_p = 3~{\rm g~cm}^{-3}$ and a radius of $s=1~{\rm mm}$ and adopting disk gas density of $\Sigma_g = 500~{\rm g~cm}^{-2}$, a disk viscosity of $\alpha_{\rm diff} \gtrsim 1.4 \times 10^{-4}$ is required for the particle to penetrate the tertiary gap via turbulent diffusion assuming a dust depletion factor of $\gamma_0/\gamma = 1000$ as in \citet{zhu12}:
\begin{eqnarray}
\label{eqn:alpha_diff}
\alpha_{\rm diff} & = & 1.4\times10^{-4} \left( \rho_p \over 3~{\rm g~cm}^{-3} \right) \left( s \over 1~{\rm mm} \right) \left( \Sigma_g \over 500~{\rm g~cm}^{-2} \right)^{-1} \nonumber \\
 & & \times \left( \ln(\gamma_0 / \gamma) \over \ln (0.001) \right)^{-1}.
\end{eqnarray}
As shown in the above equation, $\alpha_{\rm diff}$ increases inversely proportional to the gas surface density for a given solid particle density.
Therefore, when $\alpha \lesssim 1.4 \times 10^{-4}$ all the pressure bumps beyond 2~au satisfy this first condition to trap solid particles with sizes of a mm or greater.

One can do the same experiment for the second condition about the enhanced gas velocity.
Rearranging Equation (30) of \citet{zhu12}, we obtain
\begin{equation}
\alpha_{\rm egv} =  \left | \ln \left( {\Sigma_g \over \Sigma_{g,0}} \right) \right |  \left( {\mu^2 \over 9\pi} \right)^{1/4} \left( {\rho_p s \pi \over \Sigma_{g,0}} \right)^{3/4} \left( {h \over r}\right)^{-1/2},
\end{equation}
where $\mu = M_p/M_*$.
Assuming again canonical values for a mm-sized particle placed right outside of the inner tertiary gap at $\sim 2.3$~au with a $10~\%$ of gas depletion in the gap (i.e., $\Sigma_g / \Sigma_{g,0} = 0.9$), a disk viscosity of $\alpha_{\rm egv} \gtrsim 1.8 \times 10^{-5}$ is required for the particle to penetrate the tertiary gap:
\begin{eqnarray}
\label{eqn:alpha_egv}
\alpha_{\rm egv}  & = & 1.8 \times 10^{-5} \left( \left| \ln(\Sigma_g / \Sigma_{g,0}) \right | \over \left| \ln (0.9) \right | \right) \left( {\mu \over 6\times10^{-5}} \right)^{1/2}  \nonumber\\
 & & \times \left( \rho_p \over 3~{\rm g~cm}^{-3} \right)^{3/4}  \left( s \over 1~{\rm mm} \right)^{3/4} \left( \Sigma_{g,0} \over 500~{\rm g~cm}^{-2} \right)^{-3/4} \nonumber\\
 & & \times \left( {h/r \over 0.033} \right)^{-1/2}.
\end{eqnarray}

Based on Equations (\ref{eqn:alpha_diff}) and (\ref{eqn:alpha_egv}), we can infer that solid particles with sizes of $\geq 1$~mm can remain trapped in the inner primary and secondary and the outer primary and secondary pressure bumps when the disk viscosity is sufficiently low: $\alpha \lesssim 10^{-5}$. 
We note however that this is an $\alpha$ value associated with the radial mass transport only, because the analysis presented above considers only the radial movement of gas and dust.
Since disk turbulence can be anisotropic such that vertical stress is significantly stronger than the radial stress \citep[e.g.,][]{stoll17}, we caution that this value should not be taken as a representative value for the total disk turbulence.

It has to be noted that the simulation presented in this section assumes zero kinematic disk viscosity\footnote{Non-zero numerical viscosity can exist at a level of $\alpha < 10^{-5}$.}.
Using a non-zero kinematic viscosity will result in less prominent pressure bumps. 
On the other hand, the core will generate stronger spiral arms as it grows to a full Jupiter-mass and can create strong enough pressure bumps to trap solid particles even in the presence of a moderate viscosity (e.g., $\alpha \sim {\rm a~few}\times10^{-4}$; \citealt{bae17}).

Having multiple pressure bumps helps prevent solid particles from rapidly drifting inward, but their different strengths and locations in the disk can naturally explain the range of sizes and abundances of CAIs (and chondrules) seen in meteorites.
Follow-up studies, which incorporate growth of the core with more realistic disk thermal evolution and gas-dust interaction, are certainly required to further examine this scenario. 
Nevertheless, formation of multiple pressure bumps by a growing Jupiter seems to be an intriguing possibility to explain the meteoritic property in the solar system and deserves further consideration.

\section{SUMMARY AND DISCUSSION}
\label{sec:summary}

We examined how characteristics of planet-driven spiral arms, including the number of spiral arms, pitch angle of spiral arms, and arm-to-arm separation,  change as a function of disk temperature and planet mass, aiming to utilize the characteristics of observed spiral arms to constrain the mass of unseen planet and/or its position in the disk.
To summarize the findings:
\begin{enumerate}
\item A larger number of spiral arms form (1) with a smaller planet mass and (2) with a smaller $(h/r)_p$ (Figure \ref{fig:numarm}). Our parameter study shows that three or fewer spiral arms are excited interior to a planet's orbit when $M_p/M_* \gtrsim 3\times10^{-4}$ and two spiral arms when $M_p/M_* \gtrsim 3\times10^{-3}$. The number of spiral arms exterior to a planet's orbit is more sensitive to the disk temperature than the planet mass. Two outer spiral arms can excite when $(h/r)_p \lesssim 0.06$; for warmer disks only one outer spiral arm launches.
\item Using the pitch angle of spiral arms to constrain the mass of unseen planet requires accurate disk gas temperature measurements. However, because the pitch angles of planet-driven spiral arms always decrease away from the planet, this property can be a powerful diagnostic to determine whether the unseen planet is located interior or exterior to the observed spiral arms.
\item The arm-to-arm separation increases as a function of planet mass and converges to $\sim180^\circ$ for sufficiently massive planets (Figure \ref{fig:pdiff} and \ref{fig:pdiff_tot}); the exact mass depends on $(h/r)_p$. Overall, our best fit to primary-to-secondary separation $\Delta\phi_{p-s}$ and planet mass is $\Delta\phi_{p-s} = 106^\circ (q/0.001)^{0.21}$, which is in a good agreement with the relation provided by \citet{fung15}. However, we find that the arm-to-arm separation increases more steeply with the planet mass at smaller radii and for smaller $(h/r)_p$ values.
\item The relative brightness/intensity of observed spiral arms is not an ideal characteristic to determine whether a certain spiral arm is a primary or a secondary/tertiary arm because secondary/tertiary arms can create stronger perturbation than the primary arm.
\end{enumerate}

We then applied these diagnostics to the spiral arms seen in MWC~758 and Elias~2--27.
For the MWC~758 disk, it is unlikely that the recently detected point-like source excites all three spiral arms.
A more likely explanation is that the point-like source excites one of the spirals (S3 in \citealt{reggiani17}) and another yet undetected companion beyond 0."6 from the star excites the other two (S1 and S2 in \citealt{reggiani17}).
This scenario explains observed characteristics of the spiral arms reasonably well, including the radially decreasing pitch angle of the newly detected spiral arm in \citet{reggiani17} and the non-detection of this arm in previous observations at a shorter wavelength \citep{benisty15}. 
If this is the case, observations in 2018 may reveal more than $>10^\circ$ of difference in the orbital motions between S3 and the other two compared with the first epoch observation of \citet{reggiani17} taken in 2015 October.

For the Elias~2--27 disk, we emphasize that the measured constant pitch angle in \citet{perez16} is not an expected characteristic of planet-driven spiral arms, unless the actual disk temperature profile over the region spirals extend is much steeper than what is currently constrained: $T \propto r^{-1}$ or steeper vs. $T \propto r^{-0.5}$.
We conjecture that having a constant pitch angle might be a generic feature of GI-driven spiral arms.
More accurate measurements of the disk gas surface density distribution as well as theoretical developments on GI-driven spiral arms will help reveal the true nature of the two-armed spirals in the Elias~2--27 disk.

We carried out a numerical simulation to show that spiral arms driven by Jupiter's core in the solar nebula could have created multiple gaps and pressure bumps through spiral shocks.
Some of the pressure bumps can be strong enough to trap solid particles of appropriate sizes, which can help explain the extended duration of meteorite parent body accretion as well as the broad range of sizes and abundances of CAIs seen in meteorites.

The analysis presented in this paper is based on two-dimensional numerical simulations.
As recent three-dimensional simulations have shown, planet-driven spiral arms have a three-dimensional structure in a way that they are curled toward the central star and produce larger perturbations near the disk surface \citep{zhu15}.
Vertical temperature gradient can also affect the propagation of waves  \citep{lubow98,bate02,lee15}.
More thorough analysis using three-dimensional simulations is hence guaranteed, which we will address in future papers.

\appendix

\begin{deluxetable}{ccl}
\tablecolumns{3}
\tabletypesize{\small}
\tablecaption{Best Fits to $\Delta\phi_{p-s} - M_p/M_*$ relation \label{tab:fits}}
\tablewidth{0pt}
\tablehead{
\colhead{$(h/r)_p$} & 
\colhead{Radius} & 
\colhead{Best Fits} 
}
\startdata
 & $0.2~r_p$ & *$\Delta\phi_{p-s} = 135^\circ (q/0.001)^{0.29}$ \\
$(h/r)_p=0.04,$ & $0.3~r_p$ & *$\Delta\phi_{p-s} = 115^\circ (q/0.001)^{0.40}$ \\
0.05, 0.06 & $0.4~r_p$ & *$\Delta\phi_{p-s} = 120^\circ (q/0.001)^{0.30}$ \\
 & $0.5~r_p$ & *$\Delta\phi_{p-s} = 104^\circ (q/0.001)^{0.26}$ \\ 
 & $0.6~r_p$ & $\Delta\phi_{p-s} = 115^\circ (q/0.001)^{0.21}$ \\
\hline
 & $0.2~r_p$ & *$\Delta\phi_{p-s} = 91^\circ (q/0.001)^{0.20}$ \\
$(h/r)_p=$ & $0.3~r_p$ & *$\Delta\phi_{p-s} = 95^\circ (q/0.001)^{0.28}$ \\
0.07, 0.08 & $0.4~r_p$ & $\Delta\phi_{p-s} = 111^\circ (q/0.001)^{0.26}$ \\
 & $0.5~r_p$ & $\Delta\phi_{p-s} = 108^\circ (q/0.001)^{0.24}$ \\ 
 & $0.6~r_p$ & $\Delta\phi_{p-s} = 95^\circ (q/0.001)^{0.18}$ \\
\hline
 & $0.2~r_p$ & *$\Delta\phi_{p-s} = 94^\circ (q/0.001)^{0.29}$ \\
 & $0.3~r_p$ & $\Delta\phi_{p-s} = 104^\circ (q/0.001)^{0.22}$ \\
$(h/r)_p=0.1$ & $0.4~r_p$ & $\Delta\phi_{p-s} = 107^\circ (q/0.001)^{0.21}$ \\
 & $0.5~r_p$ & $\Delta\phi_{p-s} = 103^\circ (q/0.001)^{0.17}$ \\ 
 & $0.6~r_p$ & $\Delta\phi_{p-s} = 96^\circ (q/0.001)^{0.11}$ \\
\hline
 & $0.2~r_p$ & $\Delta\phi_{p-s} = 98^\circ (q/0.001)^{0.23}$ \\
 & $0.3~r_p$ & $\Delta\phi_{p-s} = 105^\circ (q/0.001)^{0.19}$ \\
$(h/r)_p=0.12$ & $0.4~r_p$ & $\Delta\phi_{p-s} = 97^\circ (q/0.001)^{0.22}$ \\
 & $0.5~r_p$ & $\Delta\phi_{p-s} = 100^\circ (q/0.001)^{0.13}$ \\ 
 & $0.6~r_p$ & $\Delta\phi_{p-s} = 107^\circ (q/0.001)^{0.05}$ \\
\hline
 & $0.2~r_p$ & $\Delta\phi_{p-s} = 112^\circ (q/0.001)^{0.20}$ \\
 & $0.3~r_p$ & $\Delta\phi_{p-s} = 113^\circ (q/0.001)^{0.14}$ \\
$(h/r)_p=0.15$ & $0.4~r_p$ & $\Delta\phi_{p-s} = 115^\circ (q/0.001)^{0.12}$ \\
 & $0.5~r_p$ & $\Delta\phi_{p-s} = 117^\circ (q/0.001)^{0.08}$ \\ 
 & $0.6~r_p$ & $\Delta\phi_{p-s} = 121^\circ (q/0.001)^{0.05}$ 
\enddata
\tablenotetext{*}{Best fits with an asterisk symbol contain arm-to-arm separations measured for the spiral arms experiencing interference/merging.}
\end{deluxetable}

\acknowledgments

We thank Lee Hartmann, Ruobing Dong, Wing-Kit Lee, and Farzana Meru for providing valuable comments on the initial draft.
We also thank anonymous referee for providing us constructive comments.
J.B. thanks Conel Alexander, Jay Melosh, Woong-Tae Kim, Kaitlin Kratter, and Kamber Schwarz for helpful conversation.
This work was supported in part by NASA grant NNX17AE31G.
Z.Z. acknowledges support from the National Aeronautics and Space Administration through the Astrophysics Theory Program with Grant No. NNX17AK40G and Sloan Research Fellowship.
We acknowledge the following:  computational resources and services provided by Advanced Research Computing at the University of Michigan, Ann Arbor; the XStream computational resource, supported by the National Science Foundation Major Research Instrumentation program (ACI-1429830); the Extreme Science and Engineering Discovery Environment (XSEDE), which is supported by National Science Foundation grant number ACI-1548562; and the NASA High-End Computing (HEC) Program through the NASA Advanced Supercomputing (NAS) Division at Ames Research Center.

\end{document}